\documentclass[]{spie}  

\usepackage{amsmath,amsfonts,amssymb}
\usepackage{graphicx}
\usepackage[colorlinks=true, allcolors=blue]{hyperref}

\usepackage{graphicx}
\graphicspath{{Figures/}}
\usepackage{amsmath}
\usepackage{amssymb}
\usepackage{algorithm}
\usepackage{algorithmicx}
\usepackage{algpseudocode}
\usepackage{bm}
\usepackage{textcomp}
\usepackage[dvipsnames]{xcolor}
\usepackage{multirow}
\usepackage[normalem]{ulem}
\usepackage{caption}
\usepackage{subcaption}
\def\eg{{\it e.g., }}

\newcommand{\etc}{{etc.}}              

\def\be{\begin{equation}}
\def\ee{\end{equation}}
\def\ba{\begin{array}}
\def\ea{\end{array}}
\def\ban{\begin{eqnarray}}
\def\ean{\end{eqnarray}}
\everymath{\displaystyle}

\usepackage[utf8]{inputenc}
\usepackage[english]{babel}
\usepackage{hyperref}

\title{Breast shape estimation and correction in CESM biopsy}

\author[]{Rubén Sánchez De La Rosa}
\author[]{Clément Jailin}
\author[]{Ann-Katherine Carton}
\author[]{Pablo Milioni}
\author[]{Laurence Casteignau}
\author[]{Serge Muller}

\affil[]{GE Healthcare, 78530 Buc, France}

\authorinfo{R.S.: E-mail: ruben.sanchezdelarosa@ge.com}

\begin{document} 
\maketitle

\begin{abstract}

\textbf{Description of purpose} Contrast-enhanced spectral mammography can be used to guide needle biopsies. However, in vertical approach the compressed breast is deformed generating a so-called bump in the paddle aperture, which may interfere with the visibility of contrast-uptakes. Local thickness estimation would provide an enhanced image quality of the recombined image, increasing the visibility of the contrast-uptakes to be targeted during the biopsy procedure. In this work we propose a method to estimate the shape of the breast bump in biopsy vertical approach. 

\textbf{Materials \& Methods} Our method consists on two steps: first, we compute a raw thickness which does not take into account the presence of contrast-uptakes; second, we use a physical model to separate the sparse iodine texture from the breast shape. This physical model is composed by a sum of Fourier components, describing the main shape of the bump, a series of low-order polynomials, describing the main compressed thickness, paddle tilt and deflection, and non-linear components describing the translation and rotation of the paddle aperture. A 3D object mimicking a bump was fabricated to test the pertinence of our shape model. Also, clinical images of 21 patients which followed CESM-guided biopsy were visually assessed. 

\textbf{Results} Comparison between raw and final estimated thickness of our 3D test object shows an error standard deviation of 0.37~mm similar to the noise standard deviation equals to 0.32~mm. The visual assessment of clinical cases showed that the thickness correction removes the superimposed low-frequency pattern due to non-uniform thickness of the bump, improving the identification of the lesion to be targeted. 

\textbf{Conclusion} The proposed method for thickness estimation is adapted to CESM-guided biopsies in vertical approach and it improves the identification of the contrast-uptakes that need to be targeted during the procedure.

\end{abstract}


\section{INTRODUCTION}

When a clinician detects suspicious findings in contrast-enhanced spectral mammography (CESM), a CESM-guided biopsy may be ordered to further investigate a sample of the suspicious tissue\cite{dromain2011dual}. Depending on the lesion location, either a lateral or vertical biopsy approach is selected~\cite{nakamura2010stereotactic}. 

In the vertical biopsy approach, the needle is introduced in the breast via a rectangular aperture in the compression paddle and the trajectory towards the suspicious lesion is approximately perpendicular to the detector. As a result of the compression load, the breast is deformed in the paddle aperture, generating a local breast deformation that is hereafter referred as bump. The height of the bump can reach up to 20~mm and the non-uniform increased breast thickness across the aperture highly affects the signal intensity of the low-and high-energy X-ray projections used to compute the recombined iodine image.

In CESM-guided biopsy, three inputs are needed to deliver contrast-enhanced mammography images: a low-energy and a high-energy image, as well as the local thickness for each pixel to be recombined~\cite{puong2007dual}. If the bump is not taken into account in the recombination algorithm, low-frequency artifacts with similar intensity as an iodine contrast-uptake appear in the recombined images. This may drastically reduce the visibility of iodine uptake patterns An accurate thickness estimation and compensation procedure for the biopsy bump is hence required.

Breast shape estimation from (log-processed) X-ray projection is a well known and studied topic \cite{burch1995method} which can be used in various applications (thickness equalization~\cite{stefanoyiannis2000digital,snoeren2005thickness,lee2020convolutional}, tomosynthesis reconstruction procedure~\cite{zhi2016artifacts,michielsen2019patient}, volumetric breast density~\cite{zhou2001computerized} \etc). 
When considering a simple linear X-ray projection model, with a non-diffracting monochromatic source, the (log-processed) mammography image results from the integral on the X-ray path of the crossed material absorption. For a single phase sample, this intensity is hence directly related to the crossed length. The difficulty lies in processing samples composed of several phases: adipose, fibro-glandular, lesions, \etc. Three main categories are found in the literature to extract the shape of the compressed breast:
\begin{itemize}
    \item \textbf{Filtering methods} (\eg, with the use of low-pass filters)~\cite{pisano2000image,stefanoyiannis2000digital,snoeren2005thickness}.
    The main issue in those techniques hat are not driven by a physical model is that it may not easily separate the shape from the texture.
    \item \textbf{Parametric models} estimation is a commonly used technique. Model of different complexity can be used~\cite{mawdsley2009accurate,snoeren2004thickness,de2014mammographic}.
    In general, those model-based techniques do not model the large complexity of the breast shape and are designed for the entire breast (not for local breast swellings).
    \item \textbf{Deep-Learning techniques} have been used for thickness equalization~\cite{lee2020convolutional,michielsen2020estimating}. Although machine learning approaches have been shown to be extremely efficient in medical image processing, they require large databases with ground truth that are not often accessible.
\end{itemize}

In addition to the image information, it can be noted that the shape extraction can be enriched with additional external measurements \eg cameras with stereo-reconstruction techniques\cite{rodriguez2017compressed,pinto2020compressed}, photogrammetry\cite{tyson2009measurement}. However, those approaches that require additional sensors lead to changes in the acquisition protocol.

In this study, a physical-based model is developed allowing to determine   the thickness profile of the breast bump. The model was validated on realistic homogeneous test objects and its performance was evaluated on clinical images.

\section{Material \& Methods}
\label{sec:MM}


\subsection{Model of the bump shape}
To accurately estimate the breast bump shape, we first estimate the breast thickness through the recombination of the low- and high-energy images, as described by Puong et al.~\cite{puong2007dual}. 

A subset of the recombined biopsy image corresponds to the biopsy hole (rectangular area of size $L_x \times L_y$ pixels), written $I(\bm x)$ with $\bm x=[x,y]$ such as  $x \in [0,L_x]$ and $y \in [0,L_y]$. This image can be written as the sum of two positive quantities: a bump shape written $\Phi(\bm x)$ and a sparse textured iodine-equivalent image that contains the potential contrast uptakes written $\Psi(\bm x)$  
\begin{equation}
    I(\bm x)=\Phi(x)+\Psi(\bm x).
\end{equation} 
The goal of the procedure is to separate the sparse iodine texture $\Psi(\bm x)$ from the breast shape $\Phi(x)$. Without additional assumptions on $\Psi(\bm x)$ and $\Phi(\bm x)$ the problem is ill-posed, in the sense that its solution is not unique. Some additional constraints have to be prescribed.
Because no simple frequency properties can be extracted from $\Psi(\bm x)$ and $\Phi(x)$ (the contrast uptake can take a large variety of size and brightness), it is here proposed to design a physical model of the bump shape.

The shape of the bump is modeled as a linear combination of motion corrected shapes:
\begin{equation}
    \Phi(\bm x, \bm \alpha, \bm u) = \sum_k^{N_k} \alpha_k A_k(\bm x+\bm u(\bm x)),
\end{equation}
with $A_k$ being a library of $N_k$ bump shapes to be designed, $\bm \alpha$ are scalar weights and $\bm u(\bm x)$ a motion correction that allows adjusting the shape positioning (\eg translations, rotation, \etc).

The bump shapes library is composed of two main contributions: (1) a simple model of the bump shape and (2) Legendre polynomials that can consider additional effects such as paddle height, tilt and deflection.

\paragraph{Bump deflection model} 
The stress in the breast when compressed leads to the apparition of the bump. While a finite element simulation of the breast deformation during compression is an interesting approach\cite{mira2018simulation}, it remains a complex procedure. A bio-mechanical model of the breast would have require the integration of a complex behavior, with parameters to be identified and a high computation time.
 We chose then describe this phenomenon by the kinematics of a thin plate under a uniform hydrostatic pressure~\cite{timoshenko1959theory}. The model was hence based on the Navier solution of a deformed plate, which is written as an infinite sum of Fourier components. 
A truncated series of $N_m=N_x\cdot N_y$ modes $\mathcal{M}_{ij}$ can be introduced 
\begin{equation}
   \mathcal{M}_{ij}(\bar{\bm x}) =  \sin{\left[ (\bar x+1)(i-1/2)\pi \right] } \cdot \sin{\left[ (\bar y+1)(j-1/2)\pi \right] },
\end{equation}
with $i\in [0,N_x]$ and $j\in [0,N_y]$ and $\bar{\bm x}=[\bar x, \bar y]$ such as  $\bar x = 2x/L_x-1$ and $\bar y = 2y/L_y-1$ are centered and normalized coordinates.

\paragraph{Shape correction terms:} The previous shapes do not consider the general thickness of the breast such as its height and possible tilts of the compression paddle. It was decided to enrich the projection basis by adding low order bivariate polynomial terms. In order to reduce the coupling between the different terms and have a well conditioned problem, Legendre polynomials can be introduced. The 1 dimensional polynomials are written
\begin{eqnarray}
L_{0}(\bar x)=1, \hspace{0.7cm}
L_{1}(\bar x)=\bar x,  \hspace{0.7cm}
L_{2}(\bar x)=\dfrac{1}{2}(3\bar x^{2}-1), \\
L_{3}(\bar x)=\dfrac{1}{2}(5\bar x^{3}-3\bar x),  \hspace{0.7cm}
L_{4}(\bar x)=\dfrac{1}{8}(35\bar x^{4}-30\bar x^{2}+3)
\end{eqnarray}
and are used to design the 2D polynomials. The set of 2D polynomials at order $k \in [0,N_p]$ are written:
\begin{equation}
    P_k^i=\left\{ 
    i \in [0,k] \ | \
    L_i(x)L_{k-i}(y)
    \right\}
\end{equation}
Because Legendre polynomials are not coupled, their amplitudes can be physically interpreted individually. Order 0 is a constant term that corrects the mean height of the compression paddle. The first order 2D polynomials correspond to the two tilts of the compression paddle (parallel and perpendicular to the chestwall). The second and third polynomials correspond to a bending of the biopsy window.

\paragraph{Resolution}
The bump shapes library is hence composed of the bump model, completed with the 2D polynomials: $\bm A(\bm x)=\{ i\in [0,N_x], j\in [0,N_y],k \in [0,N_p]  \ | \ \mathcal{M}_{ij}(\bm x) ; \bm P_k(\bm x)      \}$
The identification approach aims to minimize, over a region restricted by a mask $\omega(\bm x)$ the residual value $\Gamma(\bm x,\bm \alpha,\bm u)  =  I(\bm x) - \Phi(\bm x,\bm \alpha,\bm u)$, with $\bm \alpha$ and $\bm u$ respectively the shape and motion terms
\begin{equation}
    \underset{\bm \alpha, \bm u}{\text{Argmin}} \int_{\bm x} \omega(\bm x) \Gamma(\bm x,\bm \alpha,\bm u) ^2 d\bm x
\end{equation}
The term $\omega(\bm x)$ is a binary spatial mask that aims removing non-shape intensity (\eg foreign objects, large contrast uptakes, \etc). 

For the model of the bump, $N_p=3$, $N_x=3$ and $N_y=3$ are selected. The motion correction $\bm u(\bm x)$ corresponds to 2d rigid body motions (two translations and one rotation). The resolution of this non-linear problem is performed with a Gauss Newton algorithm. At the first iteration, the mask $\omega(\bm x)$ is initialized with 1 (considering the integration on all the biopsy window), then it is updated from a simple threshold on the residual map $\Gamma$.

\subsection{CESM-biopsy datasets}

\paragraph{Development of a test object mimicking breast deformation in compression paddle aperture}
To evaluate and validate the breast shape estimation method described in Section~\ref{sec:MM}, a new test object with known shape, mimicking the local breast deformation in the compression paddle aperture, was created. The new test object is hereafter referred to as the breast bump test object.The 3D shape of the test object was derived from images of a disposable latex glove filled with sunflower oil. The glove was imaged with SerenaBright, the CESM-guided biopsy application of GE Healthcare (\emph{GE Healthcare, Chicago, Illinois, United States}), using the vertical approach. The glove was positioned on the breast support and deformed with a biopsy compression paddle containing an aperture while applying a typical clinical compression force. Then, ten dual-energy acquisitions were made using AOP-like acquisition parameters. Projected thickness images of the glove were obtained through dual-energy recombination. To obtain a 3D representation of the glove thickness in the compression paddle aperture, the average of the ten thickness images  was backprojected using the nominal system geometry. The average of the ten thickness images allowed to reduce the noise content in the thickness image thus ensuring a smooth thickness image. The 3D glove thickness representation was then converted to \emph{.stl} file format and a 3D test object was printed  using fused deposition modeling (FDM) 3D printing technology using ABSplus P430 stratasys (ABS) printing material. Printing was performed with Stratasys 1200es (\emph{Stratasys, Rehovot, Israel}) employing 100\% infill percentage and 0.25~mm layer thickness. The FDM printing technology in combination with ABS material were chosen because the print-outs have x-ray attenuation characteristics that are very similar to breast adipose tissue, both for the low-energy and high-energy beam quality used for CESM biopsy acquisitions\cite{ivanov2018suitability}. This allowed to compute the thickness image with high accuracy.

\paragraph{Datasets to evaluate the breast shape estimation}
To evaluate the breast shape estimation method described in Section~\ref{sec:MM}, the breast bump test-object was imaged with SerenaBright. The experimental set-up is shown in Figure~\ref{fig:exp_res}. The breast bump test object was positioned on top of four cm thick sections of BR0 (\emph{CIRS, Norfolk, VA}) and it was aligned with the compression paddle aperture. This setup was imaged in AOP acquisition mode. The raw thickness map was then computed from the low- and high-energy images and the raw thickness map was compared to the ground truth thickness, composed of the breast bump test object and the 4~cm thick BR0 section.


\paragraph{Clinical dataset}
In addition, the performance of our method was evaluated on clinical images acquired with the SerenaBright application using the vertical approach. CESM-guided biopsy images from 21 patients were retrospectively collected at Hospital Del Mar, Barcelona, Spain. The set of images for each patient was composed by three targeting images (X-ray tube placed at 0, -15, and +15 degrees), and two pre-fire images (X-ray tube placed at -15 and +15 degrees). In the targeting images the biopsy needle is not present in the field of view, while in the pre-fire images the biopsy needle has been already introduced in the breast and, therefore, it can be seen during image evaluation. A visual assessment was performed on CESM recombined images obtained with the algorithm described by Puong et al. \cite{puong2007dual}, with and without local thickness correction.

\section{Results}

\paragraph{Results on test object}

\begin{figure}[htbp]
    \centering
    \begin{subfigure}[c]{0.24\textwidth}
        \includegraphics[width=\textwidth]{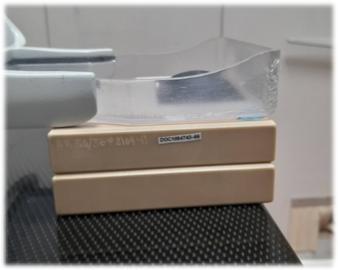}
    \end{subfigure}
    \begin{subfigure}[c]{0.75\textwidth}
        \includegraphics[width=\textwidth]{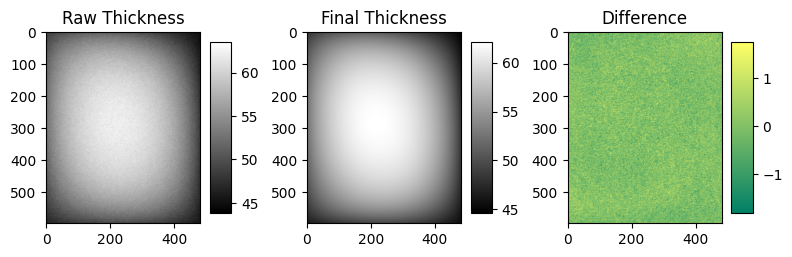}
    \end{subfigure}
    
    \vspace{-0.25cm}
    \small{\ \hspace{0.19\textwidth} [pix] \hspace{0.19\textwidth} [pix]  \hspace{0.19\textwidth} [pix]}\\
    
    \small{(1) \hspace{0.2\textwidth} (2)\hspace{0.2\textwidth} (3)\hspace{0.2\textwidth} (4)}
    \caption{(1) Image of the CESM set-up used to evaluate our shape model: the fabricated test object was placed over 50~mm thick CIRS plates of 50\% glandular equivalent tissue, (2) resulting raw thickness inside the aperture, (3) final estimated thickness using shape model inside the aperture, and (4) their difference with values are given in $mm$.}
    \label{fig:exp_res} 
\end{figure}

Figure~\ref{fig:exp_res} shows an image of the 3D printed test object mimicking local breast deformation in the aperture paddle. The maximum thickness of this test object is 16~mm.
Figure~\ref{fig:exp_res} also shows the raw thickness profile of the 3D printed test object, its final estimated thickness using the algorithm described in this work, and the difference between both. The raw thickness inside the aperture region presents an average signal of 55.5~mm (with minimum and maximum values equals to 43.3~mm and 63.4~mm respectively) and a standard deviation of 4.52~mm. The initial root-mean-square error (RMSE) of our iterative optimization algorithm is equal to 2696~mm, while the final RMSE is equal to 0.003~mm. We found that the minimum and maximum values equals to -1.8~mm and 1.7~mm respectively and a standard deviation of 0.37~mm. This last value can be compared with the standard deviation of the noise on a flat area outside the aperture, where the average thickness is 46.4~mm, that is equal to 0.32~mm. This highlights that the model is able to correctly correct the shape of the biopsy swelling.

\paragraph{Results on clinical data}

\begin{figure}[hbtp]
    \centering
    \begin{subfigure}[c]{0.28\textwidth}
        \includegraphics[width=\textwidth]{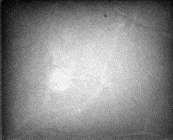}
    \end{subfigure}
    \begin{subfigure}[c]{0.71\textwidth}
        \includegraphics[width=\textwidth]{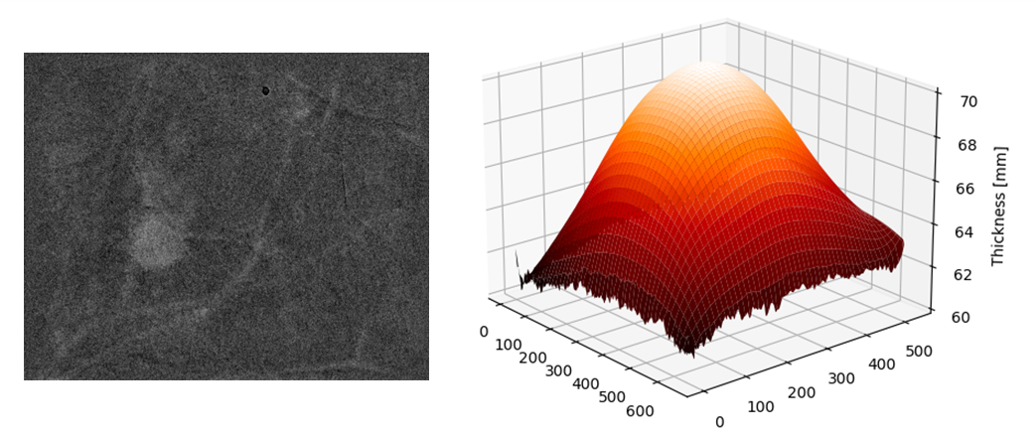}
    \end{subfigure}
    \caption{Clinical recombined images, without (left) and with (middle) applying the thickness correction, of a breast with an iodine-enhancing mass; estimated thickness of the breast bump (right).}
    \label{fig:clinical_res} 
\end{figure}

\begin{figure}[hbtp]
    \centering
    \begin{subfigure}[c]{0.28\textwidth}
        \includegraphics[width=\textwidth]{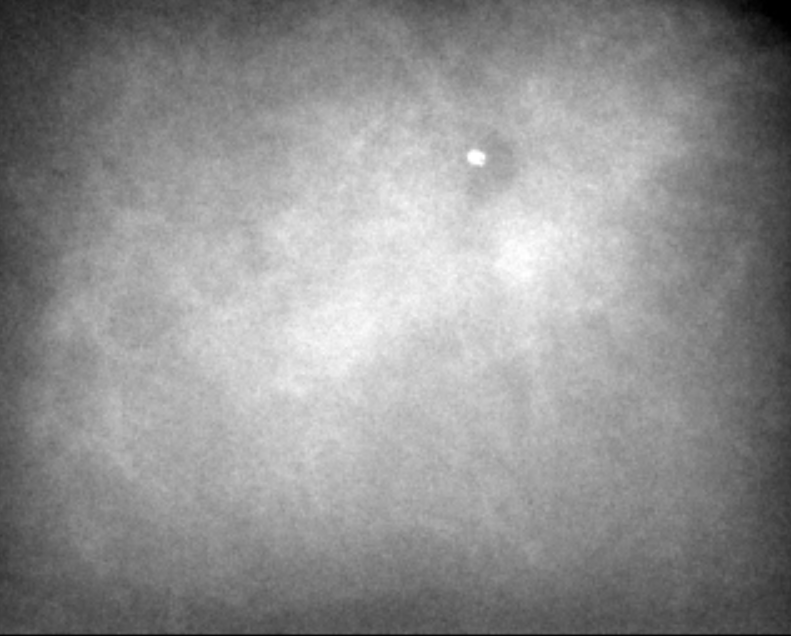}
    \end{subfigure}
    \begin{subfigure}[c]{0.70\textwidth}
        \includegraphics[width=\textwidth]{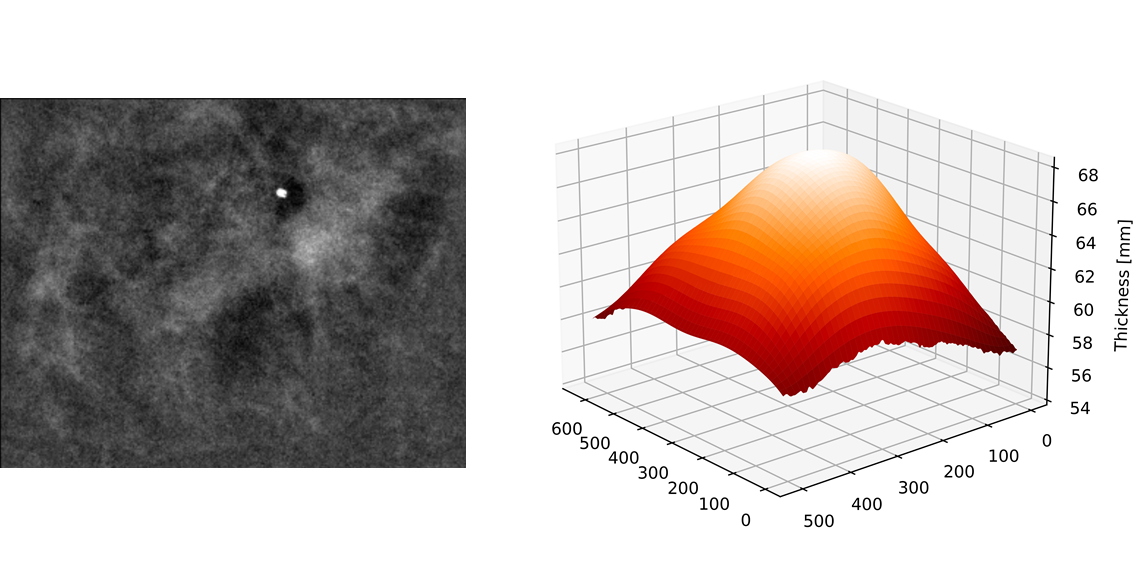}
    \end{subfigure}
    \caption{Clinical recombined images, without (left) and with (middle) applying the thickness correction, of a breast and a big contrast-uptake present in the field of view; estimated thickness of the breast bump (right).}
    \label{fig:clinical_res_bigfind} 
\end{figure}

\begin{figure}[hbtp]
    \centering
    \begin{subfigure}[c]{0.28\textwidth}
        \includegraphics[width=\textwidth]{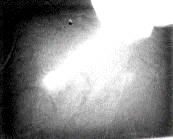}
    \end{subfigure}
    \begin{subfigure}[c]{0.70\textwidth}
        \includegraphics[width=\textwidth]{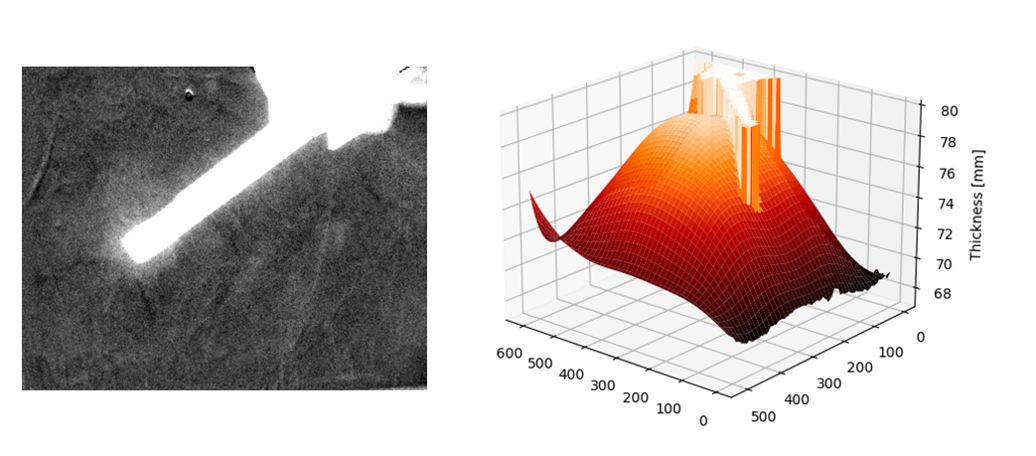}
    \end{subfigure}
    \caption{Clinical recombined images, without (left) and with (middle) applying the thickness correction, of a breast and a highly attenuating biopsy needle present in the field of view; estimated thickness of the breast bump (right).}
    \label{fig:clinical_res_needle} 
\end{figure}

Figure~\ref{fig:clinical_res} shows an example of clinical recombined images, with and without applying the thickness correction, of a breast with an iodine-enhancing mass. The thickness correction removes the superimposed low-frequency pattern due to the non-uniform thickness of the breast bump, improving the overall image quality and the identification of lesion to be targeted by the biopsy tool.
Figure~\ref{fig:clinical_res_bigfind} shows a another example of clinical recombined images, with and without applying the thickness correction, with a larger iodine-enhancing non-mass. Similarly to the previous example, the thickness correction allows to remove the low-frequency pattern due to the non-uniform thickness of the breast bump, which helps the identification of the true extent of the iodine-enhancing non-mass.
Figure~\ref{fig:clinical_res_needle} shows a third example of clinical recombined images, with and without applying the thickness correction, of a breast and a highly attenuating biopsy needle present in the field of view. The strong halo artifact present in the recombined image without thickness correction was not present in the recombined image with thickness correction, improving the overall image quality of the recombined image.  Similar results were obtained in the remaining clinical images from the 21 patients.

\section{Conclusion}
Our method aims to describe a physical model of local breast deformation induced by the biopsy breast compression paddle aperture which may reduce the visibility of iodine uptake patterns. A shape model is designed from the deflection of a thin plate under uniform pressure.
The method was tested on physical uniform phantoms and on clinical images.
We found that our model adapts well to the shape of our breast bump test object, offering an error standard deviation (0.37~mm) similar to the standard deviation due to the system noise (0.32~mm).

\textbf{New or breakthrough work to be presented:}
The proposed approach is a new thickness estimation procedure based on a physical interpretation of the geometry. It allows extracting the breast thickness without affecting, contrary to other image processing approaches, the iodine intensity of the contrast uptakes.
This procedure is robust to the presence of foreign objects (needles, wires, etc.), as well as the presence of large iodine-enhancing non-mass lesions which may occupy a large portion of the aperture region.


\section*{Acknowledgement}
The authors gratefully acknowledge Dr. Rodrigo Alcantara, Hospital del Mar - Barcelona, Spain, for the clinical images.

\section*{Compliance with Ethical Standards}
This research study was conducted retrospectively using anonymized human subject data made available by research partners. All clinical data were acquired with the approval of the ethical committee. Applicable law and standards of ethic have been respected.

\bibliographystyle{spiebib} 
\bibliography{Biblio}

\end{document}